\documentclass{tcibook}
\usepackage{fancyhea}
\usepackage{work}
\usepackage{bm}       %    enables bold math symbols  e.g.  \bm{\gamma}
\usepackage{graphicx}
\usepackage{multirow}
\usepackage{lineno}
\usepackage{threeparttable}
\usepackage[normalem]{ulem}
\usepackage{color}
\usepackage[colorlinks = true,
            linkcolor = blue,
            urlcolor  = blue,
            citecolor = blue,
            anchorcolor = blue]{hyperref}

\def\beq{\begin{equation}}
\def\eeq#1{\label{#1}\end{equation}}
\def\eeqn{\end{equation}}

\newenvironment{Eqnarray}%
   {\arraycolsep 0.14em\begin{eqnarray}}{\end{eqnarray}}
\def\beqa{\begin{Eqnarray}}
\def\eeqa#1{\label{#1}\end{Eqnarray}}
\def\eeqan{\end{Eqnarray}}

\let\bar=\overbar

\def\lsim{\mathrel{\raise.3ex\hbox{$<$\kern-.75em\lower1ex\hbox{$\sim$}}}}
\def\gsim{\mathrel{\raise.3ex\hbox{$>$\kern-.75em\lower1ex\hbox{$\sim$}}}}

\def\del{\partial}
\def\Dslash{\not{\hbox{\kern-4pt $D$}}}
\def\dslash{\not{\hbox{\kern-2pt $\del$}}}
\def\pslash{\not{\hbox{\kern-2pt $p$}}}
\def\ETmiss{\not{\hbox{\kern-4pt $E$}}_T}

\def\Dlr{\mathrel{\raise1.5ex\hbox{$\leftrightarrow$\kern-1em\lower1.5ex\hbox{$D$}}}}

\def\MSB{{\bar{M \kern -2pt S}}}
\def\msb{{\bar{\scriptsize M \kern -1pt S}}}

\def\drb{{\bar{\scriptsize D \kern -1pt R}}}

\def\authorlist#1#2{
    \vskip 0.4in
\begin{center}\begin{large} {\bf  #1 } \end{large}
    \vskip 0.2in
              #2
     \vskip 0.2in
   \end{center}
}

\begin{document}

%%  uncomment this line to use line numbers in drafts:
% \linenumbers

\pagenumbering{roman}

\parindent=0pt
\parskip=8pt
\setlength{\evensidemargin}{0pt}
\setlength{\oddsidemargin}{0pt}
\setlength{\marginparsep}{0.0in}
\setlength{\marginparwidth}{0.0in}
\marginparpush=0pt

% The content begins here

\pagenumbering{arabic}

\renewcommand{\chapname}{chap:intro_}
\renewcommand{\chapterdir}{.}
\renewcommand{\arraystretch}{1.25}
\addtolength{\arraycolsep}{-3pt}

% \thispagestyle{empty}
% \begin{centering}
% \mbox{\null}
% \rightline{\begin{tabular}{l}
% FERMILAB-CONF-xx\\
% SLAC-PUB-xx\\
%  \end{tabular}}
% \vfill

% {\Huge\bf The Future of US Particle Physics}

% \vskip 0.6in

% {\LARGE \bf Report of the 2021  US  Community Study  \\
%      on the Future of Particle Physics

%                   \smallskip

%        organized by the  APS Division of Particles and Fields}

% \vfill

% \input Frontmatter/mainauthorlist.tex

% \vfill

% \end{centering}

% \newpage
% \thispagestyle{empty}

% \mbox{\null}

% \newpage

%\pagenumbering{roman}
% \input Frontmatter/Foreword.tex 

% \newpage
% \thispagestyle{empty}

% \mbox{\null}

% \newpage

% \input Frontmatter/ExecutiveSummary.tex

% \newpage
% \thispagestyle{empty}

% \tableofcontents

% \newpage
%  \pagenumbering{arabic}

% \input Frontmatter/Summary.tex

% \newpage
% \thispagestyle{empty}

% \mbox{\null}

%\input Computation/Computation.tex

\setcounter{chapter}{6}

%% IMPORTANT:   from this file, refer to the bibliography as              
%                                                          Computation/CompF07/bibliography.tex   
%%    refer to a figure   A.pdf  as    Computation/CompF07/figures/A.pdf  .

%% To compile the overleaf for just this chapter, compile the top-level
%% "SnowmassBook-Energy.tex", not this individual "Preservation.tex" file.

\chapter{Reinterpretation and Long-Term Preservation of Data and Code}

\authorlist{
    Stephen~Bailey\textsuperscript{1},
    K.~S.~Cranmer\textsuperscript{2},
    Matthew~Feickert\textsuperscript{2},
    Rob~Fine\textsuperscript{3},
    Sabine~Kraml\textsuperscript{4},
    Clemens~Lange\textsuperscript{5}
}
{(and contributors from the community)}

\noindent
\textsuperscript{1} Lawrence Berkeley National Lab, USA \\
\textsuperscript{2} University of Wisconsin-Madison, USA \\
\textsuperscript{3} Los Alamos National Lab, USA \\
\textsuperscript{4} LPSC Grenoble, CNRS/IN2P3, France \\
\textsuperscript{5}~Paul Scherrer Institute, Villigen, Switzerland

\textbf{Note:}
This report is based upon contributed white papers, letters of interest, and
discussions with members of the community.  The authors listed above made
specific text contributions to this report or otherwise contributed white papers
cited here and specifically opted-in to authorship. 
The full community that participated in discussions leading to this report
is broader than just this author list.

\section{Executive Summary}

Careful preservation of experimental data, simulations, and analysis products maximizes their
long-term scientific return on investment by enabling new analyses and reinterpretation of the results in the future.
Key infrastructure and technical developments needed for some high-value science targets are not in scope for the operations program of the large experiments and are often not effectively funded. Increasingly, the science goals of our projects require  contributions that span the boundaries between individual experiments and surveys, and between the theoretical and experimental communities. Furthermore, the computational requirements and technical sophistication of this work is increasing. As a result, it is imperative that the funding agencies create programs that can devote significant resources to these efforts outside of the context of the operations of individual major experiments.

In this report we summarize the current state of the field and make recommendations for the future:
\begin{enumerate}
    \item Ensure that all current and future programs (both experimental and theoretical) have a strategy
    \textit{and resources} for the long term preservation of data and analysis capabilities, \textit{including beyond the lifetime of the individual projects}.  This includes supporting career development and recognition mechanisms for those who contribute to analysis and data preservation.
    \item Invest in shared cyberinfrastructure to preserve these data and support a comprehensive analysis from various experiments, surveys, simulations, and theoretical work --- both active and completed --- in order to realize their full scientific impact. The infrastructure should support the requisite theoretical inputs and computational requirements for analysis as well as metadata and APIs to track provenance and incentivize participation.  Specifically,
    \begin{enumerate}
        \item DOE Cosmic Frontier should fund a data archive organization (analogous to those supported by NASA and NSF) to preserve Cosmic Frontier datasets and simulations, and facilitate their joint analysis across the existing multiple computing centers.
        \item US funding agencies should coordinate with international partners such as the
        CERN Open Data Portal and fund additional resources as needed to ensure that all US-supported
        projects have data and analysis preservation support, including post-operations and
        including non-collider and non-experimental programs.
    \end{enumerate}
\end{enumerate}

\section{Overview}

Long-term preservation of data, simulations, and analysis results maximizes their impact by enabling
new uses beyond what was originally conceived at the time the data were obtained and the analyses performed.
It also enables joint analyses across datasets to obtain results that would be impossible if each individual
dataset or result was kept proprietary.  Funding agencies also increasingly require data to be made public,
with proprietary data becoming the exception rather than the norm.

Preserving results in a manner that is {\it useful} does not come for free, however.
Publicly released data should follow the data management principles of
Findable, Accessible, Interoperable, and Reuseable (FAIR) as described in~\cite{FAIR-paper},
which requires attention and resources throughout the end-to-end lifecycle of data generation,
processing, analysis, preservation, and distribution.  While there is broad support for the
{\it concept} of long-term preservation of data and and analyses, there is less tangible
support for the extra work needed to do so in a pragmatically useful way.  There are challenges
at the technical level to make this easier, and at the sociological level to encourage data
and analysis preservation as part of the normal lifecycle of doing research.
This includes supporting career development and recognition mechanisms for those who
contribute to analysis and data preservation.

US funding agencies currently require Data Management Plans as part of proposals, e.g. NSF\footnote{\url{https://www.nsf.gov/bfa/dias/policy/dmp.jsp}} and DOE\footnote{\url{https://www.energy.gov/datamanagement/doe-policy-digital-research-data-management}}.
These policies encourage the preservation and prompt sharing of data and analysis results, while providing
flexibility for individual cases.  e.g.~the NSF Physics policy states that
``Grantees are expected to encourage and facilitate such sharing''
while noting
``The Physics Division is not in a position to recommend a Division-specific single data sharing and archiving approach applicable to the disparate communities''.
The DOE policy requires a plan, while allowing for
``cost/benefit considerations to support whether/where the data will be preserved after direct project funding ends and any plans for the transfer of responsibilities for sharing and preservation.''

While it is good that the funding agencies allow flexibility
and do not impose a single ``one size fits all'' solution for all projects,
the currently available infrastructure for data and analysis preservation is better
suited for large collaborations that can allocate significant resources to the problem.
As will be detailed more later, the currently available resources are also dominantly
European funded or in ``HEP-adjacent'' fields such as astronomy, with comparatively little
coordination and resources for preservation of US-led HEP programs.
There is an opportunity for the field to provide better infrastructure to address the needs of the
``disparate communities'', especially smaller projects, and to provide better infrastructure
so that the ``cost/benefit considerations'' more naturally align with data/analysis preservation,
e.g.~providing cyberinfrastructure for long-term curation of data after an experiment has finished
its operations, or for preserving outputs of theoretical work that isn't tied to the data management
infrastructure of a large experiment.

Current data management plans also tend to focus on preserving the data
for the existing scientific community; there are additional opportunities to use those data
for the broader public, especially to train the next generation of scientists by expanding
access for underserved communities.

For the purposes of this chapter, we adopt the following definitions largely adapted from the contributed
white paper \cite{CompF7_WP_PreserveRecastReinterp}.

\begin{description}
    \item[1.1:] \textbf{Data}: data recorded from experiments; outputs from passing those data through
        the experiment's standard event reconstruction; and analogous simulations.
    \item[1.2:] \textbf{Derived Data}: Data that have been passed through a size reduction step that prunes
        information, but which might also add additional calculated quantities to the data files.
    \item[1.3:] \textbf{Analysis Data Products}\footnote{Originally ``Data Products'' in \cite{CompF7_WP_PreserveRecastReinterp}; we add
        ``Analysis'' here.}:
        All files containing selections of derived data and synthesised information from the various stages of an analysis.
        These might include summary plots and tables, histograms of kinematic distributions, fiducial cross sections, cross section limits, simplified model results, correlation information, analysis statistical workspace binaries or full statistical models, etc.~\cite{Cranmer:2021urp_alt}
    \item[1.4:] \textbf{Data Preservation}: The procedures, practices, and standards of ensuring the long-term      (i.e., decades beyond the end of an experiment) preservation, accessibility, and usability of data and      derived data.
    \item[1.5:] \textbf{Analysis Preservation}: The procedures, practices, and standards of ensuring the            long-term preservation, accessibility, and usability of information necessary to repeat an analysis (starting from its associated preserved data) and generate all associated Analysis Data Products.
    \item[1.6:] \textbf{Reinterpretation}: Any type of new, alternative or updated interpretation of an
        analysis or result, including the combination in, e.g., global fits or global averages. 
    \item[1.7:] \textbf{Recasting}: Reproducing the analysis logic in a simulation, considering a different
        physical process with a different phase space distribution, which might have different efficiencies and acceptances than the originally hypothesised model.
    \item[1.8:] \textbf{Program} or \textbf{Project}: an organized experiment, theoretical endeavor,
        or simulation effort that produces data outputs.
\end{description}

Note that we explicitly include simulations and theoretical work as part of ``data'', which motivates
the broader wording of ``program'' or ``project'' and not just ``experiment''.
This is the case both for simulations
generated by an experiment as part of its analyses, and simulations generated by independent research
programs, e.g.~cosmology N-body simulations or Lattice QCD calculations.
We also note that although many of the tools and prior work cited here come from collider-based
experiments and cosmology surveys, the issues discussed apply to all frontiers including dark matter,
neutrinos, lattice QCD, accelerator simulations, cosmology simulations, nuclear physics, etc.

This report draws significantly from 3 white papers contributed to the Snowmass CompF7 topic
``Reinterpretation and Long-Term Preservation of Data and Code'':
\cite{CompF7_WP_PreserveRecastReinterp} has the broadest scope, covering data and analysis preservation, and the use of those for reinterpretation and recasting of results.
\cite{CompF7_WP_Cosmo} presents issues and opportunities specific to the Cosmic
Frontier, while \cite{CompF7_WP_ADL} explores in more detail the
possibilities for Analysis Description Language(s) to promote long term preservation of analysis
logic.  This report seeks to summarize key points from those papers and provide further context,
without simply rewriting the details which are already well described in those contributed papers.

\section{Data Preservation}

Data preservation involves the long term curation, storage, and distribution of original data,
derived data, and analysis data products.  In this report we mean ``data'' in the broad sense,
including theoretical and simulation outputs, not just outputs of experimental programs.

Preserving experimental data and outputs from simulations and theory
is usually a larger scope endeavor than preserving
the data products of individual analyses, but is arguably easier to achieve since its solutions
can be addressed through the planning, budgets, and personnel resources of individual projects
and host laboratories.  A key challenge in this area is support for data preservation after
the end of the operations phase.  This still requires tangible resources, but often lacks a
specific structure for where to budget them since the original grant has finished.
This challenge is especially acute for smaller projects that
don't have the decades of generational support like those of the LHC experiments.

The CERN Open Data Portal \cite{cern-data-policy} provides infrastructure and long-term
custodial responsibility for data and software released by LHC
experiments~\cite{atlas-open-data-policy, cms-open-data-policy, lhcb-open-data-policy, alice-open-data-policy}.
It has also agreed to host non-LHC data from PHENIX and BaBar, though it is not scoped as a
general archiving resource for all HEP experiments worldwide.
While this provides the necessary infrastructure for large collider-based experiments, the community is lacking
equivalent resources in support of smaller and non-collider experiments, as well as support for
theoretical and simulation outputs.

Within astronomy, NASA and NSF support multiple data archive centers for long term curation of
public datasets, e.g.~\url{https://archive.stsci.edu}, \url{https://www.ipac.caltech.edu},
and \url{https://astroarchive.noirlab.edu}.
These preserve the data and promote their re-use beyond the lifetime of the missions and
original observing proposals that generated the data.  The DOE lacks an equivalent coordinated data archive structure for its
Cosmic Frontier experiments, leading to ad-hoc arrangements of lead labs hosting data or otherwise
delegating preservation responsibilities to other centers; see \cite{CompF7_WP_Cosmo} for further details.
Also lacking is coordination between the archive centers
and computing centers with sufficient resources to fully analyze those data,
e.g.~NERSC, ALCF, OLCF, TACC, and computing clusters at other National Labs.

Data and analysis preservation practices are not well-established in the Neutrino Frontier, perhaps owing to the lack thus far of LHC-scale experiments (as measured by longevity and data output). However, as the U.S. experimental neutrino community organizes its next decades of activity around the operation of DUNE, the ability to preserve data and re-execute analyses is well-motivated. In particular, in the near-term, it is important to consider the impact that the data (and the ability to analyze it) from currently running experiments may have for validation in the early years of the DUNE physics program. 

For example, the Short-Baseline experiments operating at Fermilab have provided the community with the most detailed characterization to date of liquid argon time projection chambers as neutrino detectors, and it would be prudent to retain the ability to measure liquid argon properties from these data sets. Similarly, MINERvA, a neutrino scattering experiment that operated at Fermilab, has driven the community’s understanding of neutrino interaction physics, and provides a comprehensive data set in an energy range highly relevant to DUNE. In recognition of this, MINERvA is undertaking a comprehensive effort to preserve its data and analysis tools \cite{CompF7_LOI_MINERvA}. The goal of this effort is to preserve the experiment’s data and simulation in a format that enables the implementation of novel analysis techniques that are developed in the next decade. The data, simulation, and in many cases, results including a large set of systematic variations will be made publicly available, along with a suite of tools to support future (re)analysis. Broadly establishing the infrastructure for, and practice of, data preservation in the Neutrino Frontier will require significant resources, and existing efforts should be used as a model for future investments in this area.

\section{Preservation of Analyses}

Although smaller in individual scope, preservation of analysis outputs is arguably more challenging because it requires broader buy-in from individual authors, and there exists more heterogeneity
in products and their archival quality.

\subsection{Preservation of Analysis Data}

HEPData\footnote{\url{https://hepdata.net}}~\cite{hepdata} provides a repository for publication-related
High-Energy Physics data, augmenting the preservation of lower-level data provided by the
CERN Open Data Portal or experiment specific sites.  HEPData focuses on relatively small ($\sim$MB) tabular
data; larger tables and non-tabular analysis data products can be archived through other sites such
as Zenodo\footnote{\url{https://zenodo.org}}, which is operated by CERN but serves the scientific
community broader than just HEP analysis data.

To a large degree, HEPData and Zenodo provide the \textit{infrastructure} needed for the preservation
of analysis data products.  The primary challenge now is the sociological adoption of using that infrastructure
to usefully archive results as a standard part of the analysis and publication lifecycle.  This sociological
challenge applies not only to the individuals doing the work, but also to hiring and promotion committees
evaluating them, i.e.~there needs to be sufficient career paths and recognition to reward effort
invested in analysis preservation for the benefit of the broader community.

\subsection{Preservation of Analysis Logic}

In principle, journal papers are intended to describe analyses and results in sufficient detail
to reproduce them and to build upon them with future work.  In practice, details are often missing
and a prose description of an analysis is insufficient for extending the work,
e.g.~by reinterpreting the results using a different physics model than originally considered.

Contributed white paper \cite{CompF7_WP_PreserveRecastReinterp} section 3
explores these issues in more detail, distinguishing between ``full preservation'' (oriented
towards capturing all of the details) and ``lightweight preservation'' (focused on pragmatic
reuse with simpler computing/software requirements).  That whitepaper lists a number of tools
to simplify and thus promote analysis preservation, though their adoption is not uniform even
within individual experiments.  Reference \cite{LHCReinterpretationForum:2020xtr} also
discusses public frameworks developed by the theory community to facilitate
reinterpretation of LHC experimental analysis results.  Further coordination across experiments and
between the experimental and theoretical communities could benefit the standardization and pragmatic
usage of these frameworks.

Analysis Description Languages (ADL) provide a complimentary
\textit{declarative} model for describing analysis details independent of the computational
framework that applies that model to data.
This approach is in contrast to traditional
\textit{imperative} code that may mix physics analysis details with computational issues such
as how to distribute the work across jobs in a batch framework.
Contributed whitepaper \cite{CompF7_WP_ADL} explores the ADL approach in more detail, as does \cite{CompF7_WP_PreserveRecastReinterp} section~3.3.
While ADLs are not a complete solution applicable to every analysis, they can be a succinct and
pragmatically useful way of describing some analyses.

\textit{Useful} analysis preservation requires significant work beyond the
original analysis.  This comes with the sociological challenge of the perception that this
extra work primarily benefits others with little incentive for the original authors.
Analysis preservation tools may gain the most traction by promoting their benefits to
individual authors using them and the collaborations that encourage their use,
rather than the altruistic benefits of helping the broader scientific community.
Well curated analysis preservation logic can streamline the reapplication of a previous
analysis to an expanded dataset, and serve as a collaboration learning resource for both
new and experienced analysts when planning future work.

Beyond the sociological challenges of analysis preservation,
there is still a learning curve even for motivated and well-intentioned authors to
know what information is most useful to preserve to maximize the benefits given limited resources.
The community as a whole is still developing these best practices, e.g.~see \cite{LHCReinterpretationForum:2020xtr}
for a discussion of lessons learned and recommendations regarding reinterpretation of LHC results.
This is an area where major projects can help lead the development of standard practices for what should
be preserved and how.

There is an increasing trend towards making scientific software open source, both at the level of
data processing and at the level of individual analyses.  Although this does not guarantee the long
term preservation and usability of the code, it is certainly better to have the code available on
a widely used publicly visible site such as GitHub, Bitbucket, or GitLab (or others), rather than on
an individual investigator's laptop.  On a pragmatic level, some universities and labs have onerous
procedures for approving the public release of previously closed-source code, while they generally have
more flexible policies towards contributing to already open-source code.  By making analysis code projects
open source from the start (even if the underlying data is still proprietary), authors can minimize the
barriers to long term preservation of their code in support of analysis preservation.

Preservation of Machine Learning (ML) models presents a particular challenge since they often depend upon
3rd party industry libraries (e.g.~TensorFlow, PyTorch, Keras) whose authors and broader communities
have different priorities and timescales for what is worth preserving.  e.g.~an ML model trained with
library ABC version X.Y may not be compatible with whatever version is available 5 or 10 years from now,
and the original version X.Y may not be installable on current operating systems. Preserving the ability
to evaluate models in software containers or encoding them in Open Neural network Exchange
(ONNX)~\cite{onnx} improves their long term preservation prospects, but this requires extra effort
from the original analysis authors, and it is not yet clear how well these will be supported on
the timescale of decades.

\section{Reinterpretation and Recasting of Analyses}

Detailed preservation of data and analyses enables their re-use beyond the scope of the original analyses.
This includes reinterpreting the results in combination with the outputs of other analyses (e.g. in global fits), reusing one or several existing analyses for testing new theoretical ideas (recasting), or reusing experimental or simulation data for a completely new analysis.
Designing and implementing datasets and analyses with this reuse in mind helps guide the pragmatic choices for where preservation effort is best spent.

Effective reinterpretation and recasting requires the preservation of both analysis data products and analyses, though the goals of the reinterpretation may require different levels of fidelity of the preservation.
For example, ATLAS has implemented full fidelity analysis reinterpretations internally using the RECAST framework \cite{Cranmer:2010hk} and fully preserved analysis workflows.
A large number of analyses is preserved within RECAST and the framework has already been used for a number of reinterpretation studies within the collaboration.
CMS has similarly implemented much more lightweight solutions, with a smaller scope focus of statistically combining analyses that explore complementary final
states.

Extensive efforts are being pursued outside of the collaborations to develop public software packages for the task of reinterpretation and/or recasting~\cite{CompF7_WP_PreserveRecastReinterp, LHCReinterpretationForum:2020xtr}.
They enable the whole HEP community to assess the impact of the experimental results beyond the interpretations provided by the experiments but heavily rely 
on the amount and quality of \textit{public} information from the experiments
(on all levels from analysis logic to analysis data products, as discussed in earlier sections).
Aiming at statistical statements about the results, e.g.~in setting limits on the parameters of new models or in performing global fits, they strongly benefit from extended information on the statistical modeling~\cite{Cranmer:2021urp_alt}. 

Public reinterpretation frameworks typically supply a database of implemented analyses. Different tools provide distinct analyses coverage and sometimes different implementations of the same analysis.
However, the proliferation of analyses and tools, and the lack of interoperability between the tools, can make the complete coverage of a physics case in reinterpretation studies difficult. 

While a unified file format for analysis implementation (a longstanding challenge) would be a significant boon for reinterpretation workflows, in the near term improvements could be made by the creation of a centralized (meta)database where the analyses available in the specific tools and the corresponding validation material can be acquired.
Additionally, creating standards and specifications for input and output formats, as well as statistical treatments would simplify adoption of the various tools.
Reinterpretation and recasting also motivates extensive exploration of the complementarities between collider results and other experimental results with global fits.
These fits expose limitations in coverage by the experiments and can identify which un(der)covered physics searches are most viable.

Robust reinterpretation and reuse encourages applying reproducible principles early in the analysis design and creating high quality analysis data products, as well as the FAIR-ification of code and analysis data products from (theory) reinterpretation studies outside the experimental collaborations at the same level as experimental analyses.

\section{Equity, Diversity, and Inclusion}

Well curated data opens significant opportunities for Equity, Diversity, and Inclusion (EDI), in that
it grants access to world-class scientific datasets for individuals and institutions that lack the
resources to participate directly in the original experiments.
e.g.~the Sloan Digital Sky Survey (SDSS) Voyages program \cite{SDSS_Voyages} provides tools and lesson
plans to use the public SDSS data in high school and undergraduate education.
Similarly, the DESI High project\footnote{\url{https://github.com/michaelJwilson/desihigh/}} provides
a high school level introduction to Dark Energy using early data from the
Dark Energy Spectroscopic Instrument.  DESI High has been used for multiple outreach programs to
minority-serving high schools.  ATLAS data on the CERN Open Data Portal have been used for
masterclasses for high school students\footnote{\url{https://opendata.cern.ch/docs/about-atlas}} and 
the ATLAS Open Data Initiative\footnote{\url{http://opendata.atlas.cern}} provides course material
for physics undergraduate and masters students.
All of these projects provide the documentation and computing resources
necessary to do the programs (not just the data), thus lowering barriers to entry.

It is important to note that simply making the data available isn't sufficient to achieve EDI goals.
In addition to providing data access, the scientific community must build upon that with non-expert documentation,
curricula, mentorship, and access to computing resources to leverage those results for under-served populations.

\section{Challenges and Opportunities}

We conclude by summarizing some challenges and opportunities for reinterpretation and long-term
preservation of data and code.

Opportunity: there is broad support for the concept of data and analysis preservation, i.e.~few
people think that it is a bad idea or substantially drains community resources from other more deserving work.

Challenge: at the individual level, it does take more work to make analysis results usefully preservable
for the future, and there is a perception that this work brings limited benefit for the original authors
and primarily benefits others who aren't doing the work.  To promote analysis preservation, the community
needs to tangibly incentivize those doing the work in a way that benefits their career.

Opportunity: journal papers already aspire to describe results in sufficient detail to reproduce them,
and there is broad community buy-in that this is a good thing.
Computing technologies such as data archives, software containers, open source software, Jupyter notebooks, and analysis description languages enable achieving that common goal with a higher level of fidelity.

Challenge: funded infrastructure for preservation beyond the lifetime of individual projects is limited.
Who has custodial responsibility of data and analysis results after the completion of a project,
and who decides what is worth preserving?  The community needs a system with a low barrier to
long-term curated preservation of data and results, without it becoming a prohibitively expensive
dumping ground for data that are never used again.

In summary, there exists a significant opportunity to maximize the return on investment
through long term preservation of data and analyses, enabling their reuse for additional
scientific results and broadening access for traditionally underrepresented groups.
Services such as the CERN Open Data Portal, Zenodo, and HEPData (all European funded)
are a good starting point, but US funding agencies should expand support to ensure that
smaller and non-collider projects (including theory) have a similar level of long term
archival support, especially beyond the operations phase of individual programs and
including coordination with the computing resources necessary to use the archived
data and analyses.

%%%%%%%%%%%%%%%%%%%%%%%%%%%%%%%%%%%%%%%%%%

%  If you would like to use BibTEX for the bibliography, please feel free to do so.  It is not required.

%  To use BibTeX,

%    1.  uncomment the following two lines,
%    2.  comment out everything below from  \begin{thebibliography}{99}   to \end{thebibliography).
%    3.  create the file  myreferences.bib in this directory, and process this file in the usual way

%\bibliographystyle{JHEP}
%\bibliography{Computation/CompF07/myreferences} 

%%%%%%%%%%%%%%%%%%%%%%%%%%%%%%%%%%%%%%%%%

%\begin{thebibliography}{99}
%\input Computation/CompF07/bibliography.tex
%\end{thebibliography}

% Bibliography
%\clearpage
\bibliographystyle{JHEP}
\bibliography{Computation/CompF07/preservation,Computation/CompF07/reinterpretation,Computation/CompF07/extra}

\end{document}